\documentclass[sigconf,authorversion]{acmart}
\AtBeginDocument{%
  \providecommand\BibTeX{{%
    \normalfont B\kern-0.5em{\scshape i\kern-0.25em b}\kern-0.8em\TeX}}}

\copyrightyear{2022}
\acmYear{2022}
\setcopyright{acmlicensed}
\acmPrice{15.00}
\acmDOI{10.1145/3510003.3510099} 
\acmISBN{978-1-4503-9221-1/22/05}
\acmConference[ICSE 2022]{The 44th International Conference on Software Engineering}{May 21–29, 2022}{Pittsburgh, PA, USA}

\usepackage{algorithmic}
\usepackage[linesnumbered,ruled,vlined]{algorithm2e}
\usepackage{graphicx}
\usepackage{textcomp}
\usepackage{xcolor}
\usepackage{color}
\usepackage{multirow}
\usepackage{subfigure}
\usepackage{CJKutf8}
\usepackage{wrapfig}
\usepackage{hyperref}
\usepackage{mathrsfs}
\usepackage{siunitx}
\usepackage{adjustbox}
\usepackage{booktabs}
\usepackage{bchart}
\usepackage{balance}

\usepackage{color,colortbl}
\usepackage{comment}
\definecolor{Gray}{gray}{0.9}
\definecolor{darkgreen}{rgb}{0,0.5,0}

\definecolor{gray91}{RGB}{91 128 184}
\definecolor{gray81}{RGB}{192 80 77}
\definecolor{gray71}{RGB}{245 195 66}
\definecolor{gray61}{RGB}{179 154 204}
\definecolor{gray51}{RGB}{186 218 84}

\usepackage[framemethod=TikZ]{mdframed}
\makeatletter
\@namedef{ver@lineno.sty}{9999/12/31}
\@namedef{opt@lineno.sty}{}
\makeatother
\usepackage[frozencache,cachedir=.]{minted}
\usepackage{tikz}
\usepackage{pgfplots}
\usepackage{enumitem}

\tikzset{
  chart/.style={
    legend label/.style={font={\scriptsize},anchor=west,align=left},
    legend box/.style={rectangle, draw, minimum size=5pt},
    axis/.style={black,semithick,->},
    axis label/.style={anchor=east,font={\tiny}},
  },
  bar chart/.style={
    chart,
    bar width/.code={
        \pgfmathparse{##1/2}
        \global\let\bar@w\pgfmathresult
    },
    bar/.style={very thick, draw=white},
    bar label/.style={font={\bfseries\small},anchor=north},
    bar value/.style={font={\footnotesize}},
    bar width=.75,
  },
  pie chart/.style={
    chart,
    slice/.style={line cap=round, line join=round, very thick,draw=white},
    pie title/.style={font={\bf}},
    slice type/.style n args={3}{
        ##1/.style={fill=##2},
        values of ##1/.style={}
    }
}}
\pgfdeclarelayer{background}
\pgfdeclarelayer{foreground}
\pgfsetlayers{background,main,foreground}

\newcommand{\pie}[3][]{
    \begin{scope}[#1]
    \pgfmathsetmacro{\curA}{90}
    \pgfmathsetmacro{\r}{1}
    \def\c{(0,0)}
    \node[pie title] at (90:1.3) {#2};
    \foreach \v/\s in{#3}{
        \pgfmathsetmacro{\deltaA}{\v/100*360}
        \pgfmathsetmacro{\nextA}{\curA + \deltaA}
        \pgfmathsetmacro{\midA}{(\curA+\nextA)/2}

        \path[slice,\s] \c
            -- +(\curA:\r)
            arc (\curA:\nextA:\r)
            -- cycle;
        \pgfmathsetmacro{\d}{max((\deltaA * -(.5/50) + 1) , .5)}

        \begin{pgfonlayer}{foreground}
        \path \c -- node[pos=\d,pie values,values of \s]{$\v\%$} +(\midA:\r);
        \end{pgfonlayer}

        \global\let\curA\nextA
    }
    \end{scope}
}

\tikzset{DNA Style/.style={minimum size=1.0cm, draw=black, line width=1pt, inner sep = 3pt}}{}

\newcounter{ColumnCounter}

\newcommand*{\PreviousNode}{}%
\newcommand*{\DNASequence}[2][Mark]{%
    \def\Sequence{#2}%
    \def\PreviousNode{}%
    \foreach [count=\xi] \Label/\Color in \Sequence {%
        \IfStrEq{\Color}{}{\def\Color{white}}{}
        \edef\NodeName{#1-\arabic{ColumnCounter}}
        \IfStrEq{\PreviousNode}{}{%
            \node [DNA Style, fill=\Color, anchor=west] (\NodeName) {\Label};
            \xdef\PreviousNode{\NodeName}%
        }{
            \node [DNA Style, fill=\Color, anchor=west, xshift=-\pgflinewidth] at (\PreviousNode.east)(\NodeName) {\Label};
            \xdef\PreviousNode{\NodeName}%
        }
        \stepcounter{ColumnCounter}
    }
}%

\definecolor{bblue}{HTML}{4F81BD}
\definecolor{rred}{HTML}{C0504D}
\definecolor{ggreen}{HTML}{9BBB59}
\definecolor{ppurple}{HTML}{9F4C7C}

\usepackage{pifont}
\newcommand{\cmark}{\ding{52}}%
\newcommand{\xmark}{\ding{53}}%

\hypersetup{colorlinks,
      linkcolor=ACMPurple,
      citecolor=ACMPurple,
      urlcolor=ACMDarkBlue,
      filecolor=ACMDarkBlue}
      
\newcommand{\todoviolet}[1]{\textcolor{violet}{~#1}}
\newcommand{\mygreen}[1]{\textcolor{blue}{~#1}}

\newcommand{\mytodocpurple}[1]{\textcolor{purple}{~#1}}

\newcommand{\etal}{\hbox{\emph{et al.}}\xspace}
\newcommand{\eg}{\hbox{\emph{e.g.}}\xspace}
\newcommand{\ie}{\hbox{\emph{i.e.}}\xspace}

\newcommand{\scc}[1]{\todoviolet{[scc: #1]}}
\newcommand{\civi}[1]{\mygreen{[civi: #1]}}

\newcommand{\bella}[1]{\mytodocpurple{[bella: #1]}}

\newcommand*{\mycode}{\fontfamily{lmtt}\selectfont}

\def\BibTeX{{\rm B\kern-.05em{\sc i\kern-.025em b}\kern-.08em
    T\kern-.1667em\lower.7ex\hbox{E}\kern-.125emX}}

\mdfdefinestyle{MyFrame}{%
    linecolor=black,
    outerlinewidth=0.3pt,
    roundcorner=5pt,
    skipabove = 4pt,
    innertopmargin=5pt, 
    innerbottommargin=5pt, 
    innerrightmargin=5pt,
    innerleftmargin=5pt,
    backgroundcolor=gray!10!white
    }

\begin{document}

\title{DeepFD: Automated Fault Diagnosis and Localization for Deep Learning Programs}


\author{Jialun Cao}
\affiliation{
  \institution{The Hong Kong University of Science and Technology, and
   Guangzhou \\HKUST Fok Ying Tung Research Institute}
  \city{Hong Kong}
  \country{China}
}
\email{jcaoap@cse.ust.hk}

\author{Meiziniu Li}
\orcid{0000-0001-5947-4030}
\affiliation{%
  \institution{The Hong Kong University of Science and Technology
  }
  \city{Hong Kong}
  \country{China}
}
\email{mlick@cse.ust.hk}

\author{Xiao Chen}
\affiliation{%
  \institution{Huazhong University of Science and Technology}
  \city{Wuhan}
  \country{China}
}
\email{xchencr@hust.edu.cn}

\author{Ming Wen*}
\affiliation{%
  \institution{Huazhong University of Science and Technology}
  \city{Wuhan}
  \country{China}
}
\email{mwenaa@hust.edu.cn}

\author{Yongqiang Tian}
\orcid{0000-0003-1644-2965}
\affiliation{%
    \institution{University of Waterloo, Canada, and The Hong Kong University of Science and Technology, China}
  \city{Waterloo}
  \country{Canada}
}
\email{yongqiang.tian@uwaterloo.ca}

\author{Bo Wu}
\affiliation{%
  \institution{MIT-IBM Watson AI Lab}
  \city{Cambridge, MA}
  \country{U.S.}
}
\email{bo.wu@ibm.com}

\author{Shing-Chi Cheung*}
\thanks{* Corresponding author.}
\affiliation{
  \institution{The Hong Kong University of Science and Technology, and
   Guangzhou \\HKUST Fok Ying Tung Research Institute}
  \city{Hong Kong}
  \country{China}
}
\email{scc@cse.ust.hk}


\renewcommand{\shortauthors}{CAO, et al.}


\begin{abstract}
As Deep Learning (DL) systems are widely deployed for mission-critical applications, debugging such systems becomes essential. 
Most existing works identify and repair suspicious neurons on the trained Deep Neural Network (DNN), which, unfortunately, might be a detour.
Specifically, several existing studies have reported that many unsatisfactory behaviors are actually originated from the faults residing in DL programs. 
Besides, locating faulty neurons is not actionable for developers, while locating  the faulty statements in DL programs can provide developers with more useful information for debugging.
Though a few recent studies were proposed to pinpoint the faulty statements in  DL programs or the training settings (\eg too large learning rate), they were mainly designed based on predefined rules, leading to many false alarms or false negatives, especially when the faults are beyond their capabilities. 

In view of these limitations, in this paper, we proposed DeepFD, a learning-based fault diagnosis and localization framework which maps the fault localization task to a learning problem.
In particular, it infers the suspicious fault types via monitoring the runtime features extracted during DNN model training, and then locates the diagnosed faults in DL programs. It overcomes the limitations by identifying the root causes of faults in DL programs instead of neurons, and diagnosing the faults by a learning approach instead of a set of hard-coded rules.
{
The evaluation exhibits the potential of DeepFD. It correctly diagnoses 52\% faulty DL programs, compared with around half (27\%) achieved by the best state-of-the-art works. Besides, for fault localization, DeepFD also outperforms the existing works, correctly locating 42\% faulty programs, which almost doubles the best result (23\%) achieved by the existing works.}

\end{abstract}

\copyrightyear{2022} 
\acmYear{2022} 
\setcopyright{acmlicensed}\acmConference[ICSE '22]{44th International Conference on Software Engineering}{May 21--29, 2022}{Pittsburgh, PA, USA}
\acmBooktitle{44th International Conference on Software Engineering (ICSE '22), May 21--29, 2022, Pittsburgh, PA, USA}
\acmPrice{15.00}
\acmDOI{10.1145/3510003.3510099}
\acmISBN{978-1-4503-9221-1/22/05}





\maketitle



\section{Introduction}\label{sec:intro}
Deep learning (DL) software has been actively deployed for applications such as fraud detection, medical diagnosis, face recognition, and autonomous driving~\cite{bugrepair20,bugcharacter19,taxonomy20}. 
Such software comprises DL programs, which encode the structure of a Deep Neural Network (DNN) model and the process by which the model is trained. 
Various studies~\cite{CrashLocator14,DeepMutation18,DeepMutation19,MODE18,SEDL18,wardat21DeepLocalize,ammann2016introduction} have been conducted to understand and detect DL program faults.\footnote{{The existing related works may use other terms like `'bug'', ``defect'' or ``issue''. This paper uses ``fault'' as a representative of all such related terms to refer to any human-made mistake in the DL program that leads to functionally insufficient performance~\cite{taxonomy20}.}}

Yet, debugging DL programs is still challenging~\cite{bugrepair20,MODE18,wardat21DeepLocalize,tfbug18,DeepFault19}. 
Unlike conventional software programs, the decision logic based on which a DNN model makes predictions is not explicitly encoded by the DL program's control flow~\cite{DeepFault19}. 
Instead, predictions are made by propagating inputs against the tuned parameters through a trained DNN model. The program plays the crucial role by defining, guiding and monitoring the model training process (e.g., defining network structures, training strategies and hyperparameters) based on a training set to indirectly tune the parameters of the DNN model.

Due to the stochastic nature of DL computation, a DNN model is unable to make every prediction correctly, and an incorrect prediction result does not necessarily indicate a fault in the underlying program. 
Therefore, it is hard to debug DL computation faults using conventional fault localization techniques~\cite{CrashLocator14,spectrum09,debugging09,taming09,flWeb10,faultPrioritization13,Locus16,empiricalFL21} based on individual correct and incorrect prediction results.
Techniques~\cite{MODE18,DeepFault19,Apricot19}
have been proposed to debug such faults by identifying and locating suspicious neurons or layers in the trained DNN model. 
Specifically, they draw an analogy between a faulty neuron (or layer) in a DNN model and a faulty statement (or branch) in a conventional program. With the analogy, they adapt fault localization techniques such as the spectrum-based ones to detect suspicious neurons or layers based on statistical analysis. 
However, directly tuning the weights of neurons after training cannot facilitate pinpointing the flaws in DL programs. 
Even worse, after tuning, developer may ignore the faults in the program, thus leaving the real root causes (e.g., inappropriate loss function and optimizer setting) of the unsatisfactory behavior uncovered.

Recent works addressed the problem (\ie debugging DL systems) in a more practical way. 
For instance, UMLAUT~\cite{Umlaut21} was proposed to detect program faults using predefined rules, and provide {advice} on how to fix them. 
AutoTrainer~\cite{autotrainer} was proposed to detect five predefined problems that might occur during training DNN models, with {solutions}
to eradicate the detected problems. These rule-based approaches do pinpoint more accurate root causes of the faults (\ie lacking of data normalization), or provide more actionable advice on how to fix the DL program or tuning the hyper-parameters  (\ie set the learning rate between $10^{-7}$ to 0.1). However, such hard-coded rules may induce false alarms and are incapable of revealing faults that go beyond the capability of those predefined rules. 
For example, UMLAUT reports a fault whenever
the output layer is not activated by {\mycode softmax}, while in fact, the activator of the output layer varies case by case. 
Besides, the pre-defined rules can over-simplify fault detection to limited symptoms exhibited by several types of faults 
(e.g., the example presented in Section~\ref{sec:motivation}). 


In view of the above-mentioned limitations, we look for an alternative.
Specifically, we propose a learning-based fault diagnosis and localization framework, DeepFD, which maps a fault localization task to a learning problem. 
In particular, it diagnoses and locates faults in a DL program by inferring the suspicious fault types using the features extracted in the model training process. The intuition behind is that the trend or distribution of certain external or internal variables (e.g., loss or gradients) in a program's training process can suggest the likelihood of faults and their types. 
Such an intuition is also echoed by our observation that the value of variables in a DL program can follow certain patterns during the model training process, and these patterns exhibit strong correlations with certain types of faults.
For instance, if the loss variable's value oscillates wildly, the training is likely using an inappropriate learning rate~\cite{autotrainer}. 
However, it is difficult to set a specific threshold for such oscillations and identify the existence of learning rates fault accordingly. We, therefore, resort to a learning-based approach. 

However, the design of DeepFD needs to address two main challenges. 
First, there is no existing off-the-shelf dataset (i.e., faulty DL programs with the ground-truth of faults) available that is sufficient to enable the learning of fault-related features. 
Though one may seed faults into correct DL programs to construct such a training set, how to seed diverse faults effectively into the programs, and further determine whether the DNN model is indeed faulty after fault seeding remains unknown.
Second, there are few references of deploying learning algorithms to locate DL program faults while the effectiveness of a learning-based technique often requires a set of high-quality features.
However, since there are enormous parameters and outputs during DNN training, it is infeasible to store all the values during the continuous training iterations, and use such values as features.
Therefore, how to select qualified features that can be utilized for effective localization of DNN faults remains to be challenging.

\begin{listing}[t!]
\begin{minted}{python}
# Model construction
model = Sequential()
model.add(Dense(input_dim=2, 
                output_dim=4, 
                init="glorot_uniform"))
model.add(Activation("sigmoid"))
model.add(Dense(input_dim=4,
                output_dim=1,
                init="glorot_uniform"))
model.add(Activation("sigmoid"))

# Training Configuration
sgd = SGD(lr=0.05,                                            # Bug 1
          l2=0.0, decay=1e-6, 
          momentum=0.11, nesterov=True)
model.compile(loss='mean_absolute_error', optimizer=sgd)      # Bug 2
model.fit(train_data, label, nb_epoch=1000, batch_size=4)     # Bug 3
\end{minted}
\caption{A Faulty Code Snippet Extracted from StackOverflow \#31556268.}
\label{listing:motivate}
\end{listing}

To address the first challenge, we collected a benchmark with 58 real faults in the wild with patches and analyzed the prevalent fault types made by developers. As a result, five common fault types are observed. We seeded faults of these types to hundreds of DL programs from a recent benchmark~\cite{autotrainer} to generate a training set with thousands of faulty DL programs, serving for the learning part of DeepFD. Furthermore, we adopted the generalised linear model (GLM)~\cite{GLM} and Cohen's effect size d~\cite{cohen1992power} to determine whether a fault-seeded DL program is statistically different from and worse than the original program, thus determining the oracle of the seeded programs.
To address the second challenge, inspired by the runtime information used in previous studies~\cite{wardat21DeepLocalize,autotrainer,Umlaut21,rauschmayr2021amazon}, we designed and traced the runtime information such as loss, gradients, and the number of times loss increases. Then, we applied statistical analysis on each trace of runtime information, resulting in a list of extracted runtime features, which can represent the process of this DNN model training. Thereby, with the extracted features as inputs, and the seeded faults as ground-truth labels, the fault diagnosis problem can, therefore, be translated into a classification problem. 

In summary, our work makes the following major contributions:
\begin{itemize}[leftmargin=*]
    \item \textbf{\textit{Originality:}} We identify a set of features that can be used to classify major types of faults in DL programs, which are extracted from the DNN training process. 
    Leveraging these features, we propose a learning-based fault localization framework, DeepFD, for DL programs. It is able to diagnose multiple faults and identify their individual fault types. For each of these faults, it can further locate the faulty code snippet in the concerned DL program. 
    
    \item \textbf{\textit{Benchmark:}} We build a benchmark containing
    58 faulty DL programs collected from 
    Stack Overflow and GitHub. Each of them include faults, patches and line numbers of the faulty statements. 
    This benchmark can serve for future researches on DNN debugging and repair tools. 
    \item \textbf{\textit{Usefulness:}} We encapsulate the fault seeding, faulty program checking, feature extraction, fault diagnosis and localization into the DeepFD framework, and open-sourced it~\cite{DeepFD}.
    The tool is extensible for seeding diverse faults and mutation testing independently. It can also be adapted to extract more runtime features for other learning-based works. 

    \item \textbf{\textit{Evaluation:}} 
    {The evaluation exhibits the potential of DeepFD. It can correctly diagnose 52\% cases, compared with half (27\%) achieved by the best state-of-the-art works. Besides, for fault localization, DeepFD also outperforms the existing works, reaching 42\%, while the existing works range from 10\% to 23\%.}
\end{itemize}



\section{Motivation}\label{sec:motivation}

\begin{figure*}[t!]
    \centering
    \resizebox{1.0\textwidth}{!}{
    \includegraphics{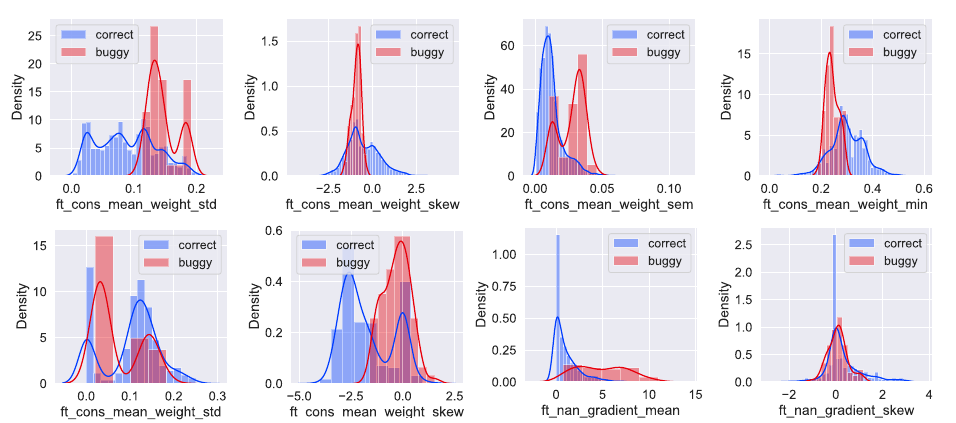}
    }
    \caption{\textbf{Correlation Between Types of Faults and Features.} The first row illustrates correlations between certain features with unsuitable \textbf{learning rate}, the second row with unsuitable \textbf{loss function}. The naming convention of features starts with `ft' (feature), followed by the name of runtime data and the applied statistical operators.}
    \label{fig:distribution}
\end{figure*}


Listing~\ref{listing:motivate} shows a buggy DL program extracted from StackOverflow.\footnote{\url{https://stackoverflow.com/questions/31556268/how-to-use-keras-for-xor.}} The program constructs a DNN model for the XOR problem, but the model's accuracy gets stuck at around 50\%. The faults in this DNN program include: (1) an inappropriate learning rate (line 13), (2) unsuitable loss function (line 16) and (3) insufficient training epoch (line 17). It is challenging for developers to diagnose and localize these faults all at once.

We applied three state-of-the-art techniques, \ie, AutoTrainer~\cite{autotrainer}, UMLUAT~\cite{Umlaut21} and DeepLocalize~\cite{wardat21DeepLocalize} to this buggy program to examine whether they can help diagnose and locate the faults. 
Note that we do not compare with automated machine learning (AutoML) algorithms~\cite{AutoML19} 
{since these works are not for debugging a faulty program, but selecting machine learning algorithms according to the user-provided data~\cite{AutoML19}}.
However, our goal is to debug a faulty program and identify the root causes. To reduce the randomness of training process, we ran each work 10 times then report the result. The results are shown in Table~\ref{tab:motivate}. 
Specifically, AutoTrainer cannot detect any faults over the 10 runs, let alone diagnosing the faults. The faults are escaped from the AutoTrainer's detection because this buggy program does not exhibit the abnormal symptoms that are predefined by AutoTrainer such as gradient vanish~\cite{GVGE91,GVGE14,GVGE19} and dying ReLU~\cite{dyingRelu20}. On the other hand, though UMLUAT and DeepLocalize are able to detect the existence of faults, the diagnosed faults are inaccurate. 
Specifically, UMLUAT falsely alarmed that the last layer has two faults (\ie missing Softmax layer before loss, and missing activation functions).
In fact, these two alarms will be fired as long as the output layer is not activated by {\mycode Softmax}. The only root cause that has been correctly diagnosed by UMLUAT is the inappropriate learning rate, but it is marked as a warning.
On the other hand, DeepLocalize reports a fault at the output gradient of layer 3 (\ie, line 10) due to numerical errors (\eg~{\mycode NaN}) occurred when propagating to this layer. Though it does locate where the symptom happens, it fails to correctly pinpoint any of the faults.


Taking a closer look, some existing works prescribe the symptoms and map them only to the existence of potential faults, but cannot identify the specific types of faults~\cite{wardat21DeepLocalize}. 
Other works predefined rules relying on various predefined thresholds, yet it is infeasible to try out all the combinations of various thresholds to find out the optimal one~\cite{autotrainer,Umlaut21}.
As such, our learning-based framework, DeepFD, is able to overcome these limitations by learning the correlations between symptoms and specific types of faults, and learning the optimal thresholds automatically. 

The secret of DeepFD is that some runtime information
exhibits significant correlations with certain types of faults. 
Specifically, Figure~\ref{fig:distribution} shows the distributions of 
several features (see Section~\ref{sec:deepfd} for more details) when the learning rate and loss function are faulty (in red) or not (in blue). 
We can observe that there are obvious statistical differences when the learning rate or loss function are set appropriately or inappropriately. 
The observation enables us to perform fault diagnosis and localization as a learning problem via leveraging the relevant stochastic runtime information of a buggy DL program. 
Finally, for the example as shown in Listing~\ref{listing:motivate}, our approach, DeepFD, can report the four types of suspicious faults in the program: the loss function (line 16), learning rate (line 13), training epoch (line 17) and activation function (line 6 and line 10). After repairing these faults accordingly, the DNN model can achieve 100\% accuracy.\footnote{Since it is a XOR problem, there is no need to consider over-fitting.}

\begin{table}[t!]
    \centering
    {\footnotesize
    \caption{\textbf{Fault Diagnosis and Localization Results.}}
    \label{tab:motivate}
    \renewcommand{\arraystretch}{1.25}
    \resizebox{1.0\linewidth}{!}{
    \begin{tabular}{l|l|l}
    \hline
    \textbf{Method} & \textbf{Fault Diagnosis} & \textbf{Fault Localization}\\ \hline
    \rowcolor{Gray}
    AutoTrainer~\cite{autotrainer} & No training problem & -- \\ 
    UMLUAT~\cite{Umlaut21}                                                   & \begin{tabular}[c]{@{}l@{}} Critical: Missing Softmax layer before loss\\ Critical: Missing activation functions\\ Warning: Last model layer has nonlinear activation\\ Warning: Learning Rate is high\\ Warning: Possible overfitting\\\end{tabular} & --                                           \\
    \rowcolor{Gray}
    DeepLocalize~\cite{wardat21DeepLocalize} & 
    \begin{tabular}[c]{@{}l@{}}
    layer-3 Error in Output Gradient\\
    Stop at 1 epoch
    \end{tabular}
    & Layer 3 \\
    DeepFD & 
    \begin{tabular}[c]{@{}l@{}}
    Fault 1: [Loss]\\
    Fault 2: [lr] \\
    Fault 3: [Epoch]\\ 
    Fault 4: [Act]\\
    \end{tabular}
    & 
    \begin{tabular}[c]{@{}l@{}}
    Lines: [16]\\
    Lines: [13] \\
    Lines: [17]\\ 
    Lines: [6, 10]\\
    \end{tabular}
    \\
    \hline
    \end{tabular}
    }
    }
    \end{table}
\section{Approach}\label{sec:overview}

\subsection{Workflow of DeepFD}\label{sec:deepfd}
\begin{figure*}[t!]
    \centering
    \resizebox{1.0\textwidth}{!}{
    \includegraphics{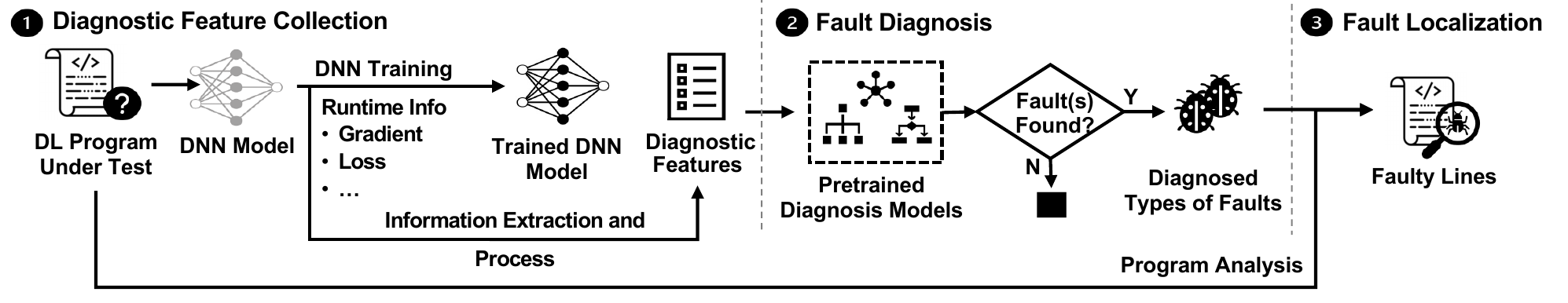}
    }
    \caption{Overview of DeepFD.
    }
    \label{fig:overview}
\end{figure*}

Figure~\ref{fig:overview} shows the workflow of DeepFD. It comprises three steps: diagnostic feature extraction, fault diagnosis and fault localization.
\\

\noindent  \textbf{Step 1: Diagnostic Feature Extraction.} 
Given a program, DeepFD constructs a DNN architecture and collects runtime data such as the loss and neurons' information by training the architecture. 
However, storing all the weights and gradients for each neuron at each training iteration is expensive.
Referencing recent works on dynamic trace collection~\cite{wardat21DeepLocalize,autotrainer} and runtime monitoring~\cite{Umlaut21,rauschmayr2021amazon}, DeepFD collects 20 runtime data (see Table~\ref{tab:runtime-info} for more details), including loss and accuracy, the number of times the loss increases, the mean or standard derivation of weights, etc. 
The data are repeatedly collected at certain intervals (e.g., every 256 batches or every epoch). 
Finally, 160 diagnostic features are extracted by applying statistical analyses (e.g., calculating the variance and skewness) to the collected data. 

\begin{table*}[thbp]
\centering
\caption{Runtime Data Collected by DeepFD During DNN Model Training}
\label{tab:runtime-info}
\renewcommand{\arraystretch}{1.0}
\resizebox{1.0\textwidth}{!}{
\begin{tabular}{l|l|l}
\hline
\textbf{Runtime Data} & \textbf{Monitored Variables} & \textbf{Description}  \\ \hline
\rowcolor{Gray}
loss & Loss & Loss on the training set. \\
acc & Accuracy & Accuracy on the training set. \\
\rowcolor{Gray}
loss\_val & Validation loss & Loss on the validation set. \\
acc\_val & Validation Aacuracy & Accuracy on the validation set. \\
\rowcolor{Gray}
nan\_loss & Loss & The number of times loss is not a number (i.e., NaN)  \\
nan\_accuracy & Accuracy & Whether there are NaN in accuracy  \\
\rowcolor{Gray}
nan\_weight & Weight & Whether there are NaN in weight  \\
nan\_gradient & Gradient & Whether there are NaN in gradient  \\
\rowcolor{Gray}
large\_weight & Weight & The number of times the maximum weight is larger than a certain threshold.  \\
decrease\_acc & Trace of accuracy & The number of times the accuracy is smaller than the last recorded accuracy.  \\
\rowcolor{Gray}
increase\_loss & Trace of loss & The number of times the loss is larger than the last recorded loss.  \\
cons\_mean\_weight & Trace of mean of weight & The number of times the mean of weight remains the same as the last record. \\
\rowcolor{Gray}
cons\_std\_weight & Trace of standard deviation of weight & Whether the standard deviation of weight remains the same as the last record.  \\
gap\_train\_test & Accuracy, Validation accuracy & Whether the gap between training accuracy and validating accuracy is too big.  \\
\rowcolor{Gray}
test\_turn\_bad & Loss, Validation loss & Whether the loss on the training set decreases while on validation set increases.  \\
slow\_converge & Trace of accuracy & Whether the accuracy of trained models is growing slowly.  \\
\rowcolor{Gray}
oscillating\_loss & Trace of Loss, Accuracy & Whether the loss is oscillating.  \\
dying\_relu & Traces of Gradient, Accuracy & \begin{tabular}[c]{@{}l@{}}There has been a set of neurons whose gradients have been 0 in the recent a few iterations and this set \\ is large forms a large portion of the whole DNN while the accuracy of the neuron network is still low.\end{tabular}  \\
\rowcolor{Gray}
gradient\_vanish & Gradient, Accuracy & Whether the gradient vanish problem occurs.  \\
gradient\_explosion & Gradient, Accuracy & Whether the gradient explode problem occurs.  \\
\bottomrule
\end{tabular}
}
\end{table*}

Table~\ref{tab:runtime-info} lists the 20 runtime data collected by DeepFD along with the monitored variables and detailed description. 
The collection of these runtime data are inspired by various existing works~\cite{wardat21DeepLocalize,autotrainer,slowConverge,osLoss18,dyingRelu20,GVGE14,GVGE14,GVGE19} 
and processed in a way following the settings used by existing studies~\cite{autotrainer,wardat21DeepLocalize}.
The runtime data are collected at predefined intervals during the training stage, resulting in 20 data sequences.
The data sequences are then processed using the eight statistical operators in Table~\ref{tab:opts} to extract diagnostic features. 
For example, the skewness of a data sequence serves as a diagnostic feature that quantifies the asymmetry of the probability distribution of the sequence with respect to its mean. After applying the statistical operators, 160 (20 * 8 = 160)
diagnostic features are extracted to characterize the training process of the given DL program. 
\\

\noindent  \textbf{Step 2: Fault Diagnosis.}
After obtaining the diagnostic features, we then infer the possible types of faults that exist in the DL program according to the features. We regard it as a multi-label classification problem, which maps the obtained features into one or more possible labels. Each label corresponds to a fault type. 
The classification relies on the predictions made by a set of pretrained diagnosis models (see Section~\ref{sec:bootstrap} for details). The diagnosis result is given by the union of the diagnosed faults predicted by each diagnosis model to maximize the number of faults diagnosed. 
Also, for each DL program under test, we run them ten times to address the randomness of DNN model training~\cite{autotrainer}.

Currently, the diagnosis models are trained to classify five major types of faults, including unsuitable loss function, optimizer, activation function, insufficient iteration and inappropriate learning rate, {which account for the majority (73.55\%) types of faults in our benchmark (see Section~\ref{sec:benchmark} for more details).}
We use these common fault types to show the promising results of how learning-based fault localization framework works, and make it extensible for supporting more types of faults.
\\

\noindent  \textbf{Step 3: Fault Localization.}
After acquiring the diagnosed types of faults, DeepFD performs fault localization at the program level. Specifically, the program is first parsed into its abstract syntax tree (AST), and DeepFD goes through the nodes of the parse tree, traverses assignment statement as well as expressions, and then identifies the places (\ie lines) where the diagnosed types of faults are defined. For example, to localize the \texttt{optimizer} in the source code, DeepFD looks for invocations to specific model training APIs (i.e., \texttt{fit}), and parses the argument and keywords of this node, and finally returns the line number where the optimizer is assigned. However, this process is not always as easy as keyword identification. For example, the \texttt{learning rate}, as a hyperparameter of the optimizer, is usually omitted in the program (\ie the default learning rate will be used), making keyword identification infeasible. In this case, DeepFD locates to the line where the optimizer is defined, while reporting the fault is in the type of learning rate, which can provide a more accurate debugging information for the developers. 
Since 
a fault may involve multiple lines, DeepFD reports a set of suspicious lines of code for each fault diagnosed.

\begin{table}[t!]
\centering
{\footnotesize
\caption{\textbf{Statistical Operators for Runtime Data Aggregation.} 
}
\label{tab:opts}
\renewcommand{\arraystretch}{1.0}
\resizebox{1.0\linewidth}{!}{
\begin{tabular}{l|l}
\hline
\textbf{Operators} & \textbf{Description} \\ \hline
\rowcolor{Gray}
max & The maximum value in a feature trace.  \\ 
min & The maximum value in a feature trace.  \\
\rowcolor{Gray}
median & The median value of a feature trace.  \\
mean & The mean value of a feature trace.  \\
\rowcolor{Gray}
var & The variance of a feature trace. \\
std & The standard deviation of a feature trace. \\
\rowcolor{Gray}
skew & The skewness of a feature trace.  \\
sem & The standard error of mean of a feature trace.  \\
\hline
\end{tabular}
}
}
\end{table}

\subsection{Diagnosis Model Construction}\label{sec:bootstrap}
Two decisions are to be made in constructing diagnosis models. First, we need to choose which machine learning algorithms to construct the models. We choose K-Nearest Neighbors~\cite{KNN,KNNIntro,KNNsurvey07,KNNSurvey08}, Decision Tree~\cite{DT,decisionTreeSurvey,rokach2005decision} and Random Forest~\cite{RandomForest,breiman2001random,ho1998random}) to construct three diagnosis models. These three algorithms are chosen because of their wide adoption, explainability and simplicity. Second, we need to decide how to aggregate their results. We choose to union their results to maximize the number of faults diagnosed. We will evaluate the individual effectiveness of the three diagnosis models in Section 5. 

\begin{figure*}[t!]
    \centering
    \resizebox{1.0\textwidth}{!}{
    \includegraphics{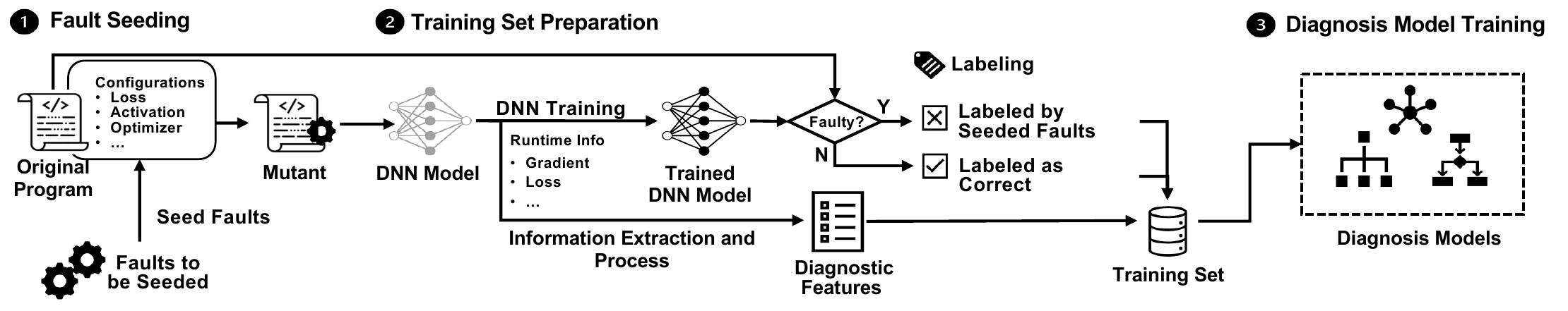}
    }
    \caption{The Workflow of Diagnosis Models Construction. 
    }
    \label{fig:model-pre}
\end{figure*}

The three diagnosis models are used in the second step (\ie fault diagnosis) of DeepFD, serving as the classifiers mapping diagnostic features into different types of faults that might exist.
The diagnosis models are constructed in three steps as shown in Figure~\ref{fig:model-pre}.
\\

\noindent \textbf{Step 1: Fault Seeding.}\label{sec:faultseeding}
This step is to prepare sufficient training samples for the diagnosis models. Since there is no off-the-shelf training set for this purpose, we are inspired by the idea of fault seeding in mutation testing~\cite{mutationSurvey11} to seed faults into normal programs.
However, fault seeding for DL programs needs to address multiple challenges. First, what types of faults should we seed? 
{Though there are several existing empirical studies~\cite{taxonomy20,bugrepair20,tfbug18,bugcharacter19}, most of the collected buggy programs are not reproducible because of incomplete or missing code snippets, unavailable training sets, or program crashes. Without reproduction, inappropriate plausible fixes may be included, inducing noise to our study.}
To address this challenge, we constructed a benchmark of 58 buggy DL programs by reproducing the failures of the DL program faults collected by recent empirical studies~\cite{bugcharacter19,bugrepair20,taxonomy20,wardat21DeepLocalize} from Github and StackOverflow. 
For each reproduced faulty program, we investigated its characteristics, including the number of faults, the fault types, as well as the code differences between the faulty version and its fixed version.
Furthermore, we found that a faulty DL program often contain faults of more than one type. Since most statements are involved in a DL program execution, the effects of multiple faults in a faulty execution can easily cascade. 
To mimic the situation, we randomly seed up to five types of faults in each program mutant. 
Second, how to seed concrete faults for a specific type of fault?
Adapting DeepCrime~\cite{deepcrime21}, we designed seven fault-seeding mutation operators in Table~\ref{tab:mut-opt} for the five fault types.

We explain the design of these operators and their differences from existing works. 
The first two mutation operators target at \textbf{loss functions}. Loss functions measure the difference between the ground truth and the predicted values, which can be further divided into probabilistic loss functions and regression ones, suiting for classification and regression tasks, respectively. 
When mutating the loss functions, DeepCrime~\cite{deepcrime21} randomly picks one from all the other available loss functions regardless of which category the loss function is.
On the contrary, we first find out the category of loss used by the given DL program, and then randomly select one from {another} category. It is because losses from another category are more likely to be unsuitable for the original task, and thus the generated mutant is more likely to be faulty. 
Next, to seed faults in the \textbf{learning rate} setting, we increase or decrease the learning rate, deliberately setting them to an extremely large or small value. While DeepCrime only consider setting learning rates to extremely small values. 
Furthermore, for the training \textbf{epochs}, we assign the epoch to a small value by randomly dividing the current number of epoch with a random number within 10 to 50, aiming at generating a small enough epoch. While the existing work only generate a random number, which may be even larger than the current number of epochs. 
Finally, for the \textbf{activation function} and \textbf{optimization function}, we randomly choose another function from the available function sets apart from the original one.
Note that though existing works~\cite{operator20,deepcrime21}
can seed more faults, their works are designed for mutation testing, which serves a different purpose from that of DeepFD. {The fault seeding step in DeepFD serves as a preparation for the diagnosis model training to perform the final fault diagnosis and localization. It is also extensible to support seeding more types of faults based on our framework.}
\\

\begin{table*}[thbp]
\centering
\caption{Seeded Faults and the Corresponding Fix Patterns}
\label{tab:mut-opt}
\renewcommand{\arraystretch}{1.15}
\resizebox{1.0\textwidth}{!}{
\begin{tabular}{l|l|l|l||l|l}
\hline
\rowcolor{Gray}
Mutation Operator & Target & Search Range & Parameter & Fix Patterns & Description \\ \hline
\begin{tabular}[c]{@{}l@{}}Change loss function to those for \\ classification problems\end{tabular} & Loss Function & Enumerate & \begin{tabular}[c]{@{}l@{}}``categorical\_crossentropy'', \\ ``sparse\_categorical\_crossentropy'',\\ ``binary\_crossentropy''\end{tabular} & \multicolumn{1}{l|}{\multirow{2}{*}{Loss Function}} & \multicolumn{1}{l}{\multirow{2}{*}{\begin{tabular}[c]{@{}l@{}}This group of fixes adjusts the loss function which helps \\ the training process to identify the deviation from the \\ learned and actual examples.\end{tabular}}} \\ \cline{1-4}
\begin{tabular}[c]{@{}l@{}}Change loss function to those for \\ regression-based problems\end{tabular} & Loss Function & Enumerate & \begin{tabular}[c]{@{}l@{}}``mean\_absolute\_error'', \\ ``mean\_absolute\_percentage\_error'', \\ ``mean\_squared\_error''\end{tabular} & \multicolumn{1}{l|}{} & \multicolumn{1}{l}{} \\ \hline
\rowcolor{Gray}
Change activation function in layers & Activation Function & Enumerate & \begin{tabular}[c]{@{}l@{}}``relu'',``sigmoid'',``softmax'',``softplus'',\\ ``softsign'',``tanh'',``selu'',``elu'',``linear''\end{tabular} & \multicolumn{1}{l|}{Activation} & \multicolumn{1}{l}{\begin{tabular}[c]{@{}l@{}}This group of fixes changes the activation function  \\ used in a layer to better match the problem.\end{tabular}} \\ \hline
Decrease number of epoch & Epoch & Range & originEpoch / random(10,50) & \multicolumn{1}{l|}{Iterations} & \multicolumn{1}{l}{\begin{tabular}[c]{@{}l@{}}This group of fixes adjusts the number \\ of times the training process will be run.\end{tabular}} \\ \hline
\rowcolor{Gray}
Change optimization function & Optimization Function & Enumerate & \begin{tabular}[c]{@{}l@{}}``SGD'', ``RMSprop'', ``Adam'', \\ ``Adadelta'', ``Adagrad''\end{tabular} & \multicolumn{1}{l|}{Optimizer} & \multicolumn{1}{l}{\begin{tabular}[c]{@{}l@{}}This group of fixes modifies the optimization algorithms \\ used by the DNN model.\end{tabular}} \\ \hline
\begin{tabular}[c]{@{}l@{}}Decrease learning rate to an extreme \\ small value\end{tabular} & Learning Rate & Range & {[}1e-16, 1e-10{]} & \multicolumn{1}{l|}{\begin{tabular}[c]{@{}l@{}}Change neural \\ architecture\end{tabular}} & \multicolumn{1}{l}{\begin{tabular}[c]{@{}l@{}}This group of fixes overhauls the design of \\ the DNN's architecture including a new set \\ of layers and hyperparameters.\end{tabular}} \\ \hline
\rowcolor{Gray}
\begin{tabular}[c]{@{}l@{}}Increase learning rate to an extreme \\ large value\end{tabular} & Learning Rate & Range & {[}1, 10{]} &
\multicolumn{1}{l|}{\begin{tabular}[c]{@{}l@{}}Change neural \\ architecture\end{tabular}} & \multicolumn{1}{l}{\begin{tabular}[c]{@{}l@{}}This group of fixes overhauls the design of \\ the DNN's architecture including a new set \\ of layers and hyperparameters.\end{tabular}} \\ \hline
\end{tabular}
}
\end{table*}


\noindent \textbf{Step 2: Training Set Preparation.}
In the previous step, a set of mutants (\ie fault-seeded programs) are generated. However, not all the mutants are necessarily faulty. 
{Due to the randomness of DNN training, slight variations of DNNs' performance~\footnote{In this paper, we use ``performance'' to refer the accuracy of loss of a DNN model, as used in the existing work\cite{zhang2020machine}} are natural. Simply considering a DNN with slightly varied performance as faulty ignores the nature of randomness, and may induce many false alarms.}
We present the criteria of faulty performance and the procedure of seeding multiple faults. 
To determine whether a mutant is faulty, we first check whether there is a significant statistical difference between the distribution of the original DNN's accuracy (\ie $A_{\mathcal{D_O}}$) and that of its mutant $A_{\mathcal{D_M}}$
as adopted the following equation~\cite{operator20,deepcrime21}:

\begin{equation}  \label{eq:kill}
\resizebox{0.95\linewidth}{!}{%
$ isKill(\mathcal{N}, \mathcal{M}, X) = 
\left\{
\begin{array}{ll}  
  true & \mathrm{if\ } \mathrm{effectSize}(A_{\mathcal{D_O}}(X), A_{\mathcal{D_M}}(X))\geq \beta \\ 
        & ~\&~\mathrm{pValue} (A_{\mathcal{D_O}}(X), A_{\mathcal{D_M}}(X)) < \alpha  \\  
  false & \\  
\end{array}  
\right\}
$
}
\end{equation}

\noindent where $\alpha$ and $\beta$ are thresholds that control the statistical significance and effect size, and $X$ represents the testing set. 
Specifically, to quantify the statistical significance and the effect size, we adopted the generalised linear model (GLM)~\cite{GLM} and Cohen's d~\cite{cohen1992power}.
If the distribution of the mutant's accuracy is statistically different from that of the original model, then we further check whether the average accuracy of the mutant is worse than its original one. If the above two requirements are satisfied at the same time, the mutant is considered as faulty, and is labeled as the types of faults that has been seeded.
If not, the mutant is regarded as non-buggy.

Furthermore, for mutants with more than one fault, another challenge is how to determine whether all these seeded faults are the root causes leading to the deteriorated performance? We address this challenge by seeding faults one by one. Specifically, after obtaining a mutant with one type of fault, we then try to seed a second fault that is independent of the first one, and use Equation~\ref{eq:kill} to check whether the second fault is successfully seeded. We repeat the above process to inject the third fault and so on. The process iterates until up to five types of faults are seeded in one original program.
Finally, we run all these generated mutants together with the original programs to collect the diagnostic features as described in the Step 1 of Section~\ref{sec:deepfd}. These diagnostic features and their labels are then used to train the diagnosis models of DeepFD. 
\\

\noindent \textbf{Step 3: Diagnosis Model Training.}
We treat the fault diagnosis as a multi-label classification problem mapping a faulty program to the multiple types of faults that it contains. We construct three diagnosis models using
the three widely-used and effective machine learning algorithms (i.e., K-Nearest Neighbors~\cite{KNN,KNNIntro,KNNsurvey07,KNNSurvey08}, Decision Tree~\cite{DT,decisionTreeSurvey,rokach2005decision} and Random Forest~\cite{RandomForest,breiman2001random,ho1998random}) to learn the correlations between diagnostic features and types of faults. 
Specifically, the K-Nearest Neighbor algorithm assumes that similar samples exist in close proximity. It clusters samples into K groups according to the distance between samples. The decision tree algorithm predicts the value of a target variable by learning simple decision rules inferred from the features. The random forest algorithm is an ensemble learning method, which operates by constructing a multitude of decision trees during training and {outputting the result that is favored by most of the decision trees.} 
Since a sample can have multiple labels, we adopt the multi-label version of these algorithms~\cite{multiKNN,multiDT,wu2016ml}, which classifies a given sample into a set of target labels. Finally, DeepFD trains three diagnosis models against the constructed training set using these algorithms. 

\section{Benchmark Construction}\label{sec:benchmark}

\begin{figure*}[t!]
    \centering
    \resizebox{1.0\textwidth}{!}{
    \includegraphics{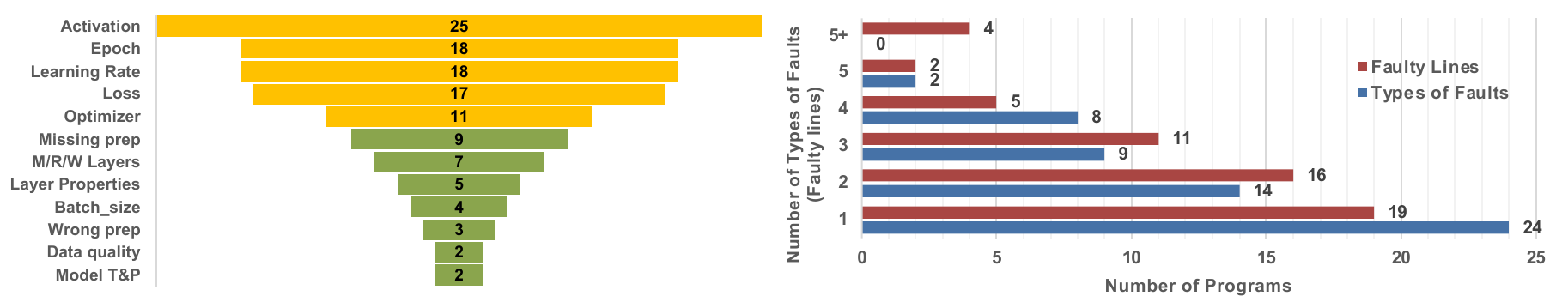}
    }
    \caption{The Statistics of Benchmark. 
    }
    \label{fig:stats}
\end{figure*}

To investigate the characteristics of faults in real faulty programs, and to evaluate the effectiveness of DeepFD, we build a benchmark with 58 buggy DL programs from StackOverflow and GitHub. The benchmark includes the artifacts required to reproduce the bugs. In this section, we elaborate how we construct the benchmark.

\textbf{Subject Collection and Selection.}
We collect the benchmark in two steps. 
First, we revisit all the benchmarks collected by previous studies~\cite{bugcharacter19,bugrepair20,taxonomy20,wardat21DeepLocalize}, e.g., 3,003 posts from StackOverflow and 2,328 commits from GitHub in total.
Besides, in order to cover the recently posted issues that are not included by the previous studies, we also search StackOverflow for recent issues following the same selection criteria as specified in~\cite{bugrepair20}. Specifically, we collect the posts from StackOverflow with accepted answers with the score greater or equal to 5, and the posted time ranges from March, 2019 till April, 2021. 
Then, we select the subject implemented by Keras and the symptom of the program is poor performance {(i.e., the program exhibits 
poor accuracy, loss during the training process)}. 
We select Keras because 46.4\% of the posts and commits concern DL programs implemented on Keras. 
We select programs with poor performance since it is the most common symptom apart from crash (i.e., the program raises an exception or crash)~\cite{tfbug18}. 
We exclude the crashed programs because they are uncommon for those written by experienced DL developers~\cite{taxonomy20}. 
For posts in StackOverflow, we exclude those without accepted answers and without source code or the training dataset since we cannot reproduce without the concerned issues.
For a similar reason, we exclude those GitHub projects that do not have training sets available. 

\textbf{Reproduction and Repair.}
We reproduce and repair the collected subjects with the following procedures. (1) We run the source code to see whether it is executable. If the source code crashes due to API upgrade, versioning or typo issues, we fix these issues; otherwise we skip this subject. (2) We examine whether the output of the source code exhibits the same symptom as described in the post. For GitHub commits, if a commit message does not describe the symptom, we capture the symptom by running the program, comparing the results before and after applying the commit. If the performance increases after applying the commit, then we take this commit into account. (3) To fix the program, we adopt the patches as suggested by the accepted answers in StackOverflow, as well as other patches recommended in other replies. 
For GitHub commits, we repair the program according to the committed changes. 
Finally, we obtain a benchmark with 58 faulty DNN programs, along with (1) the patches, usually more than one; (2) the types of faults and (3) line numbers where the faults are introduced. If a fault (e.g., adding more layers)
is not localized down to a specific piece of code, we recorded the line numbers whether the patch should be added. 
Our benchmark is made publicly available~\cite{DeepFD}.

\textbf{Statistics of the Benchmark.} 
The statistics of our benchmark are given in Figure~\ref{fig:stats}. In the left part of the Figure, we present the number of each type of faults in the benchmark. In particular, the taxonomy of faults follows the existing study~\cite{taxonomy20}, including the activation function, optimizer, loss function, hyperparameters (including the suboptimal learning rate, the suboptimal number of epochs, and the suboptimal batch size), training data quality, model type and properties, layer properties, missing/redundant/wrong layer, missing preprocess, and wrong preprocessing. The top five types of faults (\ie, activation function, suboptimal number of epoch, suboptimal learning rate, loss function and optimizer) are highlighted in yellow, accounting for the majority (73.55\%) types of faults in our benchmark. 
In addition, we also realized that one program usually contains more than one types of faults, so we further present the statistics of the number of fault types (in blue) and number of faulty lines (in red) in one program. 
Note that if one type of fault occurs in multiple places in the program (\eg, the activation function in all the layers are inappropriately set), we only count one since these faults belong to the same type. 
Similarly, one faulty line of code is counted regardless of how many faults contained in this line. {For the cases with missing layers or missing data preprocessing, we regard their number of faulty lines as one.}
From the second diagram of Figure~\ref{fig:stats}, we can see that over half of the programs in the benchmark contain more than one fault type, and most of them involve more than one line. 

\section{Evaluation}
\label{sec:experiments}
We study three research questions to evaluate the relevancy of diagnostic features and the effectivness of DeepFD in this section. 
\begin{itemize}[leftmargin=*]
\item \textbf{RQ1. Are the extracted features relevant to fault diagnosis?} 
To investigate whether the diagnostic features have high correlations with certain types of faults, we apply these features to perform fault diagnosis on the generated training sets to evaluate the relevancy of such extracted features based on the results. The higher relevancy, the more accurate the diagnosis will be.
\item \textbf{RQ2. How effective is DeepFD in fault diagnosis?} 
To evaluate the effectiveness of DeepFD on fault diagnosis, we compared DeepFD with existing works on the benchmark in terms of the number of identified and correctly diagnosed faulty cases. We also showed the diagnosis information reported by each work.
\item \textbf{RQ3. How effective is DeepFD for fault localization?} 
Accurate fault localization is an important step towards automated program repair.
So we further evaluate the effectiveness of fault localization on the benchmark and compare it with the existing works.
\end{itemize}

\subsection{Experiment Setup}
\label{sec:setup}
We implemented DeepFD in Python, and conducted experiments on a machine with Intel i7-8700K CPU and Nvidia GeForce Titan V 12GB VRAM.
For mutation generation, we run each DNN model 20 times to obtain a reliable statistical results. 
To address the randomness in DNN training, we run each experiment 10 times.
For the thresholds in DeepFD, we set $\alpha$ and $\beta$ to be 0.2 and 0.05, respectively, following the settings in~\cite{empiricalFL21}. 
{We used their default parameters in the diagnosis models. We did not fine-tune the parameters for better performance to facilitate reproducibility. In addition, parameter tuning is not the theme of this work. 
We collected runtime data using the same default interval as in prior work~\cite{autotrainer}.}
The experimental results are made available for validation~\cite{DeepFD}.

\textbf{Datasets and Original DL Programs.}
We performed our evaluation on six popular datasets: 
Circle~\cite{circle}, Blob~\cite{blob}, 
MNIST~\cite{lecun-mnist-2010}, CIFAR-10~\cite{cifar10-09}, IMDB~\cite{imdb} and Reuters~\cite{Reuters21578}. 
Circle and Blob are two datasets from 
Sklearn~\cite{sklearn} 
for classification tasks. 
MNIST is a gray-scale image dataset used for handwritten digit recognition.
CIFAR10 is a colored image dataset used for object recognition. 
IMDB is a movie review dataset for sentiment analysis. 
Reuters is a newswire dataset for document classification.
For the DL programs, we use the programs published by a recent work~\cite{autotrainer} as the original programs for mutant generation.
Specifically, the work~\cite{autotrainer} provided 495 DNN models and their training scripts of various DNN model structures (convolutional neural network, recurrent neural network and fully connected layers only) for these six datasets. 
Among them, we were able to reproduce the training of 233 DNN models. 
{Therefore, we used these 233 DNN models as the original models to generate mutants, following the process described in Section~\ref{sec:faultseeding}.
The statistics of generated mutants is shown in Table~\ref{tab:artifect} (Entry `Mutant'). For RQ1, the training set and validation set were split in a proportion of 7:3. The evaluation was conducted on the validation set.}

\textbf{Baselines.}
In the experiment, we compared DeepFD with three baselines: UMLAUT~\cite{Umlaut21}, AutoTrainer~\cite{autotrainer} and DeepLocalize~\cite{wardat21DeepLocalize}. 
Specifically, the first baseline, UMLAUT~\cite{Umlaut21}, is a heuristic-based framework providing an interface to debug faults in data preparation, model architecture and parameter tuning. 
In particular, UMLAUT allows users to check the structure of deep learning programs, model behavior during training.
After detecting faults, it provides a set of human-readable error messages and repair advice. 
The second baseline, AutoTrainer, is designed for detecting five types of training problems (\ie, Gradient Vanish, Gradient Explode, Dying ReLU, Oscillating Loss and Slow Convergence). 
We use its default parameters in our evaluation.
The third baseline, DeepLocalize, is able to identify the faulty layers when numerical error occurs. Both approaches monitor the runtime information (\eg, weight and gradient) during training, and report training problems or faulty layers once detected. 
Besides, DeepLocalize performs the fault localization using two methods: translating the code into an imperative representation manually or inserting customized callback functions. 
We used their second method for comparison because the second one achieves better performance than the first method according to their paper.
{Although these baselines do not explicitly point out the fault types they addressed, the fault types can still be mapped from the patches or faulty layers they provided to the fault types. After the mapping, the fault types appearing in the evaluation buggy programs can be covered by the baselines.}

The three baselines were not originally designed for fault localization. They do not locate faulty lines down to program code. 
To measure their effectiveness for fault localization, we adapt their diagnosis results as follows. 
For UMLAUT, we manually identify the lines of code relating to the error messages.
For AutoTrainer, though it does not explicitly locate the faults, it tries to repair when training problems happens by applying predefined solutions. 
Therefore, we regard the location of the solutions adopted by AutoTrainer as the fault it located, and manually identify the lines of code relating to the repaired artifact (\eg learning rate, layers' initialization) to the corresponding lines.
For DeepLocalize, we map the reported layer to the corresponding lines in the program. And because of the manual work that is involved in both fault diagnosis and localization stages, we do not compare the overall runtime of each stage.

\textbf{Evaluation Criteria.}
We adopt the following criteria in deciding whether a fault is successfully detected, diagnosed and localized. 
A fault is successfully {detected} if the detection result corresponds to the existence of a real fault, regardless of whether its root causes (\ie types of faults) has been identified.
A fault is successfully {diagnosed} if its root cause has been identified. 
If there are multiple types of faults, a successful fault diagnosis refers to those that can pinpoint at least one of them.
{
For fault {localization}, we examine whether any of the faulty lines is correctly located after a correct diagnosis.}

\begin{table*}[thbp]
\centering
\caption{\textbf{Effectiveness of Diagnosis Models of DeepFD on Labeled DNN Set}. KNN, DT and RF stand for the underlying algorithms (K-Nearest Neighbors, Decision Tree and Random Forest) of diagnosis models.}
\label{tab:artifect}
\renewcommand\arraystretch{1.1}
\resizebox{1.0\textwidth}{!}{
\begin{tabular}{c|cccccccc|cccccc}
\hline
 & \multicolumn{8}{c|}{Statistics} & \multicolumn{6}{c}{Accuracy and Average Runtime of Diagnosis Models
 }
 \\ \cline{2-15} 
 & \multirow{2}{*}{Origin} & \multirow{2}{*}{Mutant} & \multicolumn{5}{c}{\# Types of Faults} & \multirow{2}{*}{Time} & \multicolumn{2}{c|}{KNN} & \multicolumn{2}{c|}{DT} & \multicolumn{2}{c}{RF} \\ \cline{4-8} \cline{10-15} 
Dataset &  &  & 1 & 2 & 3 & 4 & 5 &  & Acc (\%) & \multicolumn{1}{c|}{Time (s)} & Acc (\%) & \multicolumn{1}{c|}{Time (s)} & Acc(\%) & Time (s) \\ \hline
\rowcolor{Gray}
MNIST & 78 & \multicolumn{1}{c}{1768} & \multicolumn{1}{c}{1027} & \multicolumn{1}{l}{302} & \multicolumn{1}{l}{295} & \multicolumn{1}{l}{134} & \multicolumn{1}{l}{10} & 0.60 & 69.29 & \multicolumn{1}{c|}{0.10} & 63.15 & \multicolumn{1}{c|}{0.01} & 81.58 & 0.12 \\
CIFAR-10 & 35 & 786 & 651 & 102 & 33 & 0 & 0 & 1.56 & 52.63 & \multicolumn{1}{c|}{0.05} & 53.94 & \multicolumn{1}{c|}{0.02} & 64.47 & 0.12 \\
\rowcolor{Gray}
Circle & 36 & 936 & 580 & 174 & 133 & 44 & 5 & 8.87 & 45.20 & \multicolumn{1}{c|}{0.23} & 55.50 & \multicolumn{1}{c|}{0.03} & 61.36 & 0.22 \\
Blob & 39 & 685 & 335 & 158 & 137 & 50 & 5 & 0.13 & 83.95 & \multicolumn{1}{c|}{0.06} & 79.01 & \multicolumn{1}{c|}{0.01} & 87.65 & 0.10 \\
\rowcolor{Gray}
Reuters & 32 & 175 & 138 & 22 & 3 & 12 & 0 & 36.30 & 79.69 & \multicolumn{1}{c|}{0.05} & 79.69 & \multicolumn{1}{c|}{0.01} & 81.25 & 0.09 \\
IMDB & 13 & 110 & 76 & 25 & 9 & 0 & 0 & 53.72 & \textbf{93.30} & \multicolumn{1}{c|}{0.04} & \textbf{88.80} & \multicolumn{1}{c|}{0.03} &  \textbf{93.30} & 0.09 \\ \hline
Total & 233 & 4,460 & 2,807 & 783 & 610 & 240 & 20 & Average & \textbf{70.68} & \multicolumn{1}{c|}{0.09} & \textbf{79.90} & \multicolumn{1}{c|}{0.02} & \textbf{78.27} & \multicolumn{1}{c}{0.12} \\ \hline
\end{tabular}
}
\end{table*}


\subsection{RQ1. Relevancy of Diagnostic Features}
We evaluate the relevancy of the extracted diagnostic features by showing the accuracy of the diagnosis models trained with these features. 
If the models perform well when using these features, then we regard these diagnostic features as relevant to fault diagnosis. 
The experimental results are shown in Table~\ref{tab:artifect}. 
Before analyzing the prediction result, we first demonstrate the statistics of the mutants generated in the first step of the Bootstrap stage. 
Specifically, for each dataset, we list the number of normal DNN models (Origin), the number of generated mutants (Mutant), the detailed distribution of how many mutation operators are applied to the generated mutants (column ``Mutation Operator'') and the average time for training each mutant in seconds (Time). 

On top of this training set, we further trained diagnosis models with three underlying algorithms (i.e., KNN, DT and RF are abbreviations of K-Nearest Neighbors, decision tree and random forest), and demonstrated the accuracy of these models. In the implementation, we normalized the features to better fit the K-Nearest-Neighbors algorithm. As shown in the entry of ``Accuracy and Average Runtime of Diagnosis Models'' in Table~\ref{tab:artifect}, the average accuracy of these diagnosis models range from 70.68\% to 79.90\% over different datasets.
Besides, the accuracy obtained varies cross datasets, ranging from 45.2\% to 93.3\%. Among all these datasets, the best performance of three underlying algorithms is achieved on IMDB dataset, with the accuracy varying from 88.8\% to 93.3\%. On the other hand, the accuracy on Circle and CIFAR-10 tend to be worse. 
The result may be caused by the unbalanced mutants with different numbers of faulty cases. 
The various performance of different underlying algorithms on datasets also suggest us to aggregate the diagnosis decisions by taking advantage of different diagnosis models.

\begin{mdframed}[style=MyFrame]
\textbf{Answer to RQ1}: The extracted diagnostic features have strong correlations with our targeted five fault types, which is reflected by the accuracy of diagnosis models, with an average accuracy varying from 70.68\% to 79.90\%, and up to 93.30\% for certain cases.
\end{mdframed}

\subsection{RQ2. Effectiveness of Fault Diagnosis.}
To answer RQ2, we evaluate the effectiveness of DeepFD in terms of fault diagnosis on the 52 buggy programs~\footnote{The rest six programs were omitted because they either were unable to be adapted to launch the methods, or crashed when the methods are applied.} from the benchmark. {Note that the subjects used for RQ2 and RQ3 do not overlap with those used in RQ1.}
Due to space limitation, Table~\ref{tab:fl-fd} shows partially our evaluation results, \footnote{We listed the cases that can be correctly diagnosed by at least three methods among UMLAUT, AutoTrainer, DeepLocalize and DeepFD. The complete experimental results are online available~\cite{DeepFD}}
including the diagnosis information provided by each method (column `Diagnosis Information of Different Methods'), whether they detect the existence of faults (column `Fault Detection') and whether the fault diagnosis is correct or not (column `Fault Diagnosis'). UMLAUT enumerates error messages with three severity levels (\ie, Error, Critical and Warning). 
AutoTrainer reports the training problem. DeepLocalize reports the place (\eg, layer) and the batch at which the predefined symptom occurs. DeepFD reports the diagnosed fault types. 

The `Summary' and `Overall Ratio' of Table~\ref{tab:fl-fd} summarize the results of each work on the benchmark. 
UMLAUT, DeepLocalize and DeepFD can detect 69\% to 96\% faulty cases. Although UMLAUT detects the most number (50) of faulty cases, only 14 of them are correctly diagnosed. In contrast, DeepFD is able to detect 47 faulty cases and correctly diagnosed 27 of them.

There are different reasons for inaccurate diagnosis. 
UMLAUT is designed for classification problems. It assumes that the output layer is always activated by {\mycode Softmax}. If this is not the case, it reports ``Missing Softmax layer before loss''. However, this assumption does not necessarily hold. Activation functions like {\mycode Linear} and {\mycode Sigmoid} are also frequently used to activate the output layer for classification and regression problems.
Indeed, most of the faults detected by UMLAUT are attributed to the violation of this assumption, causing many false alarms. 
{AutoTrainer only detects five training problems (\eg dying ReLU), leaving most cases without apparent symptoms escaped from its detection.}
DeepLocalize detects faults that cause numerical anomalies such as {\mycode NaN}, and reports the place where the anomalies happen. Yet, anomalies rarely occur at the point where the faults reside, making the fault diagnosis by DeepLocalize inaccurate. On the other hand, the effectiveness of DeepFD is limited by the number of fault types that can be classified by the diagnosis models. With more types of faults seeded, more faults can be detected and diagnosed, and thus the effectiveness of DeepFD can be improved.

\begin{mdframed}[style=MyFrame]
\textbf{Answer to RQ2}: DeepFD outperforms the existing works on fault diagnosis, with a fault diagnosis rate of 52\%, which almost doubles that of the best baseline  (\ie, 27\%). 
\end{mdframed}


\subsection{RQ3. Effectiveness of Fault Localization}

The `Fault Diagnosis' column in Table~\ref{tab:fl-fd} gives the evaluation result of fault localization. DeepFD outperforms other methods by correctly locating 42\% of faults,
which favorably compares to the 23\% by UMLAUT, the best baseline performer. 
The performance of AutoTrainer and DeepLocalize are relatively unsatisfactory, ranging from 10\% to 15\%.

Indeed, the performance on fault localization are not satisfactory, ranging from 10\% to 42\%. The reasons behind are mainly two folds. First, the ratio of correctly diagnosed cases are not high, which is at most 52\%. 
Second, considering the complicated types of faults and possible multiple patches, an effective localization strategy is to be explored.  
For example, suppose we know the root cause is the use of inappropriate activation, yet there are multiple activation functions in several layers.
Consequently, locating faults to a specific activation function at a specific layer is an outstanding challenge. 

\begin{mdframed}[style=MyFrame]
\textbf{Answer to RQ3}: The results show that DeepFD significantly outperforms the three baselines by correctly locating 42\% of the cases. In comparison, only 10\% to 23\% cases can be located by the baselines.
\end{mdframed}

\begin{table*}[thbp]
\caption{\textbf{Comparison on Diagnosis Information, Fault Detection, Diagnosis and Localization Between UMLAUT (UT), AutoTrainer (AT), DeepLocalize (DLoc) and DeepFD (DFD).
}}
\label{tab:fl-fd}
\renewcommand\arraystretch{1.13}
\resizebox{\textwidth}{!}{
{\huge
\begin{tabular}{lllllcccccccccccc}
\hline
\multicolumn{1}{l|}{\multirow{2}{*}{Post \#}} & \multicolumn{4}{c|}{Diagnosis Information of Different Methods}                                                                                                         & \multicolumn{4}{c|}{Fault Detection}                                                                   & \multicolumn{4}{c|}{Fault Diagnosis}                                                                  & \multicolumn{4}{c}{Fault Localization}                                                                          \\ \cline{2-17} 
\multicolumn{1}{l|}{}                         & \multicolumn{1}{c|}{UMLAUT}                                                                                                                                                                                                                                                                                                                       & \multicolumn{1}{c|}{AutoTrainer} & \multicolumn{1}{c|}{DeepLocalize}                                                                                                                     & \multicolumn{1}{c|}{DeepFD}    & UT                 & AT            & DLoc           & \multicolumn{1}{c|}{DFD} & UT                 & AT            & DLoc          & \multicolumn{1}{c|}{DFD} & UT                 & AT            & \multicolumn{1}{l}{DLoc} & \multicolumn{1}{l}{DFD} \\ \hline
\multicolumn{1}{l|}{48385830}                 & \multicolumn{1}{l|}{\begin{tabular}[c]{@{}l@{}}(1) Critical: Missing Softmax layer before loss\\ (2) Warning: Possible overfitting\end{tabular}}                                                                                                                                                                                                  & \multicolumn{1}{l|}{explode}     & \multicolumn{1}{l|}{\begin{tabular}[c]{@{}l@{}}Layer-1 Error in forward\\ Stop at epoch 1, batch 2\\ Accuracy: 0.13, loss: inf\end{tabular}}          & 
\multicolumn{1}{l|}{\begin{tabular}[c]{@{}l@{}} Fault 1: [act] (Lines: [-]) \\ Fault 2: [loss] (Lines: [57]) \end{tabular}}
& \cmark                    & \cmark                    & \cmark                    & \multicolumn{1}{c|}{\cmark}    & \cmark                    & \cmark                    & \cmark                   & \multicolumn{1}{c|}{\cmark}    & \cmark                    & \cmark                    & \xmark               & \cmark                        \\
\rowcolor{Gray}
\multicolumn{1}{l|}{55328966}                 & \multicolumn{1}{l|}{\begin{tabular}[c]{@{}l@{}}(1) Error: Input data exceeds typical limits\\ (2) Warning: Possible overfitting\\ (3) Warning: Check validation accuracy\\ (4) Critical: Missing Softmax layer before loss\\ (5) Critical: Missing activation functions\\ (6) Warning: Last model layer has nonlinear activation\end{tabular}}    & \multicolumn{1}{l|}{explode}     & \multicolumn{1}{l|}{--}                                                                                                                               & 
\multicolumn{1}{l|}{\begin{tabular}[c]{@{}l@{}} Fault 1: [opt] (Lines: [49])  \end{tabular}}
& \cmark                    & \cmark                    & \xmark                     & \multicolumn{1}{c|}{\cmark}    & \cmark                    & \cmark                    & \xmark                    & \multicolumn{1}{c|}{\cmark}    & \cmark                    & \cmark                    & \xmark                            & \cmark                        \\
\multicolumn{1}{l|}{34311586}                 & \multicolumn{1}{l|}{\begin{tabular}[c]{@{}l@{}}(1) Critical: Missing Softmax layer before loss\\ (2) Warning: Last model layer has nonlinear activation\end{tabular}}                                                                                                                                                                             & \multicolumn{1}{l|}{--}          & \multicolumn{1}{l|}{\begin{tabular}[c]{@{}l@{}}Batch 0 layer 2: Error in delta \\ Weights, terminating training\end{tabular}}                         & 
\multicolumn{1}{l|}{\begin{tabular}[c]{@{}l@{}} Fault 1: [lr] (Lines: [27])  \end{tabular}}
& \cmark                    & \xmark                     & \cmark                    & \multicolumn{1}{c|}{\cmark}    & \cmark                    & \xmark                     & \cmark                   & \multicolumn{1}{c|}{\cmark}    & \cmark                    & \xmark                     & \cmark                              & \cmark                        \\

\rowcolor{Gray}
\multicolumn{1}{l|}{50079585}                 & \multicolumn{1}{l|}{\begin{tabular}[c]{@{}l@{}}(1) Critical: Missing Softmax layer before loss\\ (2) Critical: Missing activation functions\\ (3) Warning: Last model layer has nonlinear activation\end{tabular}}                                                                                                                                & \multicolumn{1}{l|}{unstable}    & \multicolumn{1}{l|}{--}                                                                                                                               & 
\multicolumn{1}{l|}{\begin{tabular}[c]{@{}l@{}} Fault 1: [lr] (Lines: [44]) \\ Fault 2: [epoch] (Lines: [15])  \end{tabular}}
& \cmark                    & \cmark                    & \xmark                     & \multicolumn{1}{c|}{\cmark}    & \cmark                    & \cmark                    & \xmark                    & \multicolumn{1}{c|}{\cmark}    & \cmark                    & \cmark                    & \xmark                               & \cmark                        \\
\multicolumn{1}{l|}{47352366}                 & \multicolumn{1}{l|}{\begin{tabular}[c]{@{}l@{}}(1) Critical: Missing Softmax layer before loss\\ (2) Warning: Last model layer has nonlinear activation\end{tabular}}                                                                                                                                                                             & \multicolumn{1}{l|}{explode}     & \multicolumn{1}{l|}{\begin{tabular}[c]{@{}l@{}}Layer-12 Error in delta weights\\ Stop at epoch 1, batch 24\\ accuracy: 0.31, loss: 6.16\end{tabular}} & 
\multicolumn{1}{l|}{\begin{tabular}[c]{@{}l@{}} Fault 1: [opt] (Lines: [40])  \end{tabular}}
& \cmark                    & \cmark                    & \cmark                    & \multicolumn{1}{c|}{\cmark}    & \cmark                    & \cmark                    & \cmark                   & \multicolumn{1}{c|}{\xmark}     & \cmark                    & \cmark                    & \cmark                              & \xmark                         \\

\rowcolor{Gray}
\multicolumn{1}{l|}{59282996}                 & \multicolumn{1}{l|}{\begin{tabular}[c]{@{}l@{}}(1) Error: Input data exceeds typical limits\\ (2) Warning: Check validation accuracy\\ (3) Critical: Missing Softmax layer before loss\end{tabular}}                                                                                                                                              & \multicolumn{1}{l|}{unstable}    & \multicolumn{1}{l|}{--}                                                                                                                               & 
\multicolumn{1}{l|}{\begin{tabular}[c]{@{}l@{}} Fault 1: [epoch] (Lines: [309])  \end{tabular}}
& \cmark                    & \cmark                    & \xmark                     & \multicolumn{1}{c|}{\cmark}    & \cmark                    & \cmark                    & \xmark                   & \multicolumn{1}{c|}{\cmark}    & \cmark                    & \cmark                    & \xmark                               & \cmark                        \\

\multicolumn{1}{l|}{37624102}                 & \multicolumn{1}{l|}{\begin{tabular}[c]{@{}l@{}}(1) Critical: Missing Softmax layer before loss\\ (2) Critical: Missing activation functions\\ (3) Warning: Last model layer has nonlinear activation\\ (4) Error: Image data may have incorrect shape\\ (5) Warning: Learning Rate is high\\ (6) Warning: Check validation accuracy\end{tabular}} & \multicolumn{1}{l|}{unstable}    & \multicolumn{1}{l|}{\begin{tabular}[c]{@{}l@{}}Batch 0 layer 9: Error in Output \\ Gradient, terminating training\end{tabular}}                       & 
\multicolumn{1}{l|}{\begin{tabular}[c]{@{}l@{}} Fault 1: [lr] (Lines: [66]) \\ Fault 2: [Act] (Lines: [54, 56, 61, 64]) \end{tabular}}
& \cmark                    & \cmark                    & \cmark                    & \multicolumn{1}{c|}{\cmark}    & \cmark                    & \cmark                    & \xmark                    & \multicolumn{1}{c|}{\cmark}    & \xmark                     & \cmark                    & \xmark                               & \cmark                        \\

\rowcolor{Gray}
\multicolumn{1}{l|}{41600519}                 & \multicolumn{1}{l|}{\begin{tabular}[c]{@{}l@{}}(1) Error: Input data exceeds typical limits\\ (2) Critical: Missing Softmax layer before loss\\ (3) Warning: Last model layer has nonlinear activation\end{tabular}}                                                                                                                              & \multicolumn{1}{l|}{unstable}    & \multicolumn{1}{l|}{\begin{tabular}[c]{@{}l@{}}Batch 0 layer 6: Error in forward, \\ terminating training\end{tabular}}                               & 
\multicolumn{1}{l|}{\begin{tabular}[c]{@{}l@{}} Fault 1: [loss] (Lines: [32])  \end{tabular}}
& \cmark                    & \cmark                    & \cmark                    & \multicolumn{1}{c|}{\cmark}    & \xmark                     & \cmark                    & \cmark                   & \multicolumn{1}{c|}{\cmark}    & \xmark                     & \cmark                    & \cmark                              & \cmark                        \\ \hline 
\multicolumn{17}{c}{\huge \textbf{......}       } \\                                                                                          \\ \hline
\multicolumn{5}{r}{\huge \textbf{Summary}}                                                                                                            & \multicolumn{1}{|r}{\huge \textbf{50}} & \multicolumn{1}{r}{\huge 12} & \multicolumn{1}{r}{\huge 36} & \multicolumn{1}{r|}{\huge \textbf{47}}      & \multicolumn{1}{r}{\huge 14} & \multicolumn{1}{r}{\huge 10} & \multicolumn{1}{r}{\huge 7} & \multicolumn{1}{r|}{\huge \textbf{27}}     & \multicolumn{1}{r}{\huge 12} & \multicolumn{1}{r}{\huge 8} & \multicolumn{1}{r}{\huge 5}            & \multicolumn{1}{r}{\huge \textbf{22}}    \\

\multicolumn{5}{r}{\huge \textbf{Overall Ratio}}                                                                                                            & \multicolumn{1}{|r}{\huge \textbf{0.96}} & \multicolumn{1}{r}{\huge 0.23} & \multicolumn{1}{r}{\huge 0.69} & \multicolumn{1}{r|}{\huge 0.90}      & \multicolumn{1}{r}{\huge 0.27} & \multicolumn{1}{r}{\huge 0.19} & \multicolumn{1}{r}{\huge 0.13} & \multicolumn{1}{r|}{\huge \textbf{0.52}}     & \multicolumn{1}{r}{\huge 0.23} & \multicolumn{1}{r}{\huge 0.15} & \multicolumn{1}{r}{\huge 0.10}            & \multicolumn{1}{r}{\huge \textbf{0.42}}    \\


\bottomrule
\end{tabular}
}
}
\end{table*}
\section{Related Work}\label{sec:related-work}
\subsection{Debugging Deep Learning Systems}

Recently, a branch of techniques have been proposed to facilitate the debugging process of deep learning software systems.
Several works focus on the DNN models, locating suspicious neurons and correcting them by prioritizing or augmenting the training data. 
For instance, Ma~\etal~\cite{MODE18} leveraged the state differential analysis to identify the features that are responsible for the incorrect classification results, and then generate the inputs based on these features to correct the behaviours of the DNN models.
Eniser~\etal~\cite{DeepFault19} proposed DeepFault, a framework to identify the suspicious neurons in a DNN and then fix these errors by generating samples for model retraining.
Apricot~\cite{Apricot19} is a weight-adaption approach to fix DNN models.
Their intuition is that the limitation of the complete models may be addressed via adapting the weight of the compact models.
These previous studies concentrate on the the DNN models and they are not able to locate the faulty lines in DNN programs.

Besides, there are several studies that are closer to ours.
For instance,
AutoTrainer~\cite{autotrainer} proposed an automatic approach to detect and fix the training problems in DNN programs at runtime. It particularly focuses on detecting and repairing five common training problems: gradient vanish, gradient explode, dying ReLU, oscillating loss, and slow convergence. It encapsulates and automates the detecting and repairing process by dynamic monitoring.
However, it relies on predefined rules to perform the bug detection. 
Wardat~\etal~\cite{wardat21DeepLocalize} proposed DeepLocalize, the first fault localization approach for DNN models (such as incorrect learning rate or inappropriate loss function). Yet, it locates to layers where the symptom happens instead of where the fault's root cause resides. 
Amazon SageMaker Debugger~\cite{rauschmayr2021amazon} and UMLAUT~\cite{Umlaut21} both provide a set of built-in heuristics to debug faults in DNN models during training. 
The major difference between our work from these methods is that DeepFD does not rely on predefined rules, making it more flexible and adaptive for various types of faults.


\subsection{Automated Machine Learning}

Automated Machine Learning (AutoML) provides methods and processes to make machine learning available for non-Machine Learning experts. 
The user simply provides data, and the AutoML system automatically determines the approach that performs the best for a particular application~\cite{AutoML19}. 
In particular, there are three main problems in AutoML, including Hyperparameter Optimization (HPO), Neural Architecture Search (NAS), and meta-learning. HPO~\cite{HPO-15-evol,HPO-19-Orthogonal,HPO17-particalSwarm} search for methods to set optimal hyperparameters in ML algorithms automatically, thus reducing the human efforts necessary for applying ML. Yet, it is not always clear which hyperparameters of an algorithm need to be optimized, and in which ranges~\cite{AutoML19}. 
{In contrast, DeepFD differs from HPO in nature. 
DeepFD diagnoses and locates faults in a given model, while the HPO methods search an optimal model from scratch. }
NAS methods~\cite{NAS-17-RL,NAS-18-parameter-sharing,NAS-19-DNN,NAS-20-MemNAS,NAS-21-NN,NAS-survey,NAS18-Bayesian} are designed to automatically design more complex neural architectures.
Meta-Learning~\cite{meta-algorithm21,meta-pipeline18,metadata05,metafeature-17,metalearning-scalable18,metalearning09} aims to improve learning across different tasks or datasets. It can significantly improve the efficiency of HPO and NAS with the help of the transferred knowledge between tasks. 
Note that though these methods are not designed for debugging an existing program, it is potential to apply these methods, especially the HPO and NAS ones for repairing after faults are diagnosed and localized. The diagnosed information can serve as a guidance, narrowing down the parameters that need to be tuned, and the search space for NAS to explore.

\section{Threats to Validity}\label{sec:threat}
In this section, we discuss three threats that may affect the validity of our work.
First, the construction of benchmark (\eg the reproduction, root cause analysis and repair) involved manual inspection of the source code, which may be subjective. To reduce this threat, each subject was examined by three authors separately and the results were cross-validated. Decisions were made only if the three authors reached an agreement. 
Second, to prepare the training set for diagnosis models, we assume the original programs (\ie programs before seeding faults) are fault-free, yet it may not always be the truth. Though they are released~\cite{autotrainer} and guaranteed to be free from five training problems (\eg gradient vanish and dying ReLU), it is still possible there are hidden faults in the program. 
Third, to reproduce the results of existing works, we adopt the default values for the parameters, which may affect their performance and efficiency. Besides, some works need to manually adapt the programs in order to launch the debugging process, which may introduce unexpected variance from the original program.
Also, with the manual work involved, the time cost by each work is hard to evaluate, leaving the efficiency of each work incomparable.
Finally, since existing works cannot locate faults to the program, we carefully investigate their diagnosis information, and manually locate the faulty lines for fair comparisons, which may also slightly affect the comparison results.

%



\section{Conclusion}\label{sec:conclusion}
In this paper, we proposed DeepFD, a learning-based fault diagnosis and localization framework which maps the fault localization task to a learning problem. In particular, DeepFD diagnosis faults by classifying runtime features into possible types of faults (\eg inappropriate optimizer), then locates faulty lines to the program.
The evaluation shows the potential of DeepFD. Specifically, it correctly diagnoses 52\% of the cases, compared with half (27\%) achieved by the best state-of-the-art works. Besides, for fault localization, DeepFD also outperforms the existing works, correctly locating 42\% faulty cases on the benchmark,
which almost doubles the result (23\%) achieved by the best state-of-the-art work.


%

\section*{Acknowledgment}
This work was supported by the 
National Key Research and Development Program of China, No. 2019YFE0198100, 
National Natural Science Foundation of China (Grant Nos. 61932021, 62002125), 
the Hong Kong RGC/GRF (Grant No. 16207120),
Hong Kong ITF (Grant No. MHP/055/19),
Huawei PhD Fellowship,
and MSRA Collaborative Research Grant. 
The authors would also like to thank the anonymous reviewers for their comments and suggestions. 

\newpage
\balance
\bibliographystyle{ACM-Reference-Format}
\bibliography{reference}

\end{document}